\documentclass[conference]{IEEEtran}
\IEEEoverridecommandlockouts
\usepackage{amsmath,amssymb,amsfonts}
\usepackage{algorithmic}
\usepackage{graphicx}
\usepackage{textcomp}
\usepackage{xcolor}
\def\BibTeX{{\rm B\kern-.05em{\sc i\kern-.025em b}\kern-.08em
    T\kern-.1667em\lower.7ex\hbox{E}\kern-.125emX}}

\usepackage{hyperref}
\usepackage[style=ieee]{biblatex}
\usepackage{quantikz}
\usepackage{multirow}
\usepackage{orcidlink}
\usepackage{booktabs}
\makeatletter
\let\MYcaption\@makecaption
\makeatother

\usepackage[font=footnotesize]{subcaption}

\makeatletter
\let\@makecaption\MYcaption
\makeatother

\definecolor{pqc}{RGB}{150, 255, 150}
\definecolor{so2}{RGB}{255, 255, 120}
\definecolor{su2}{RGB}{255, 200, 150}

\begin{document}

\title{Multi-class quantum classifiers with tensor\\network circuits for quantum phase recognition}

\author{\IEEEauthorblockN{Marco Lazzarin}
\IEEEauthorblockA{\textit{CNR-IFN,}
Milan, Italy}
\IEEEauthorblockA{\textit{Università degli Studi di Milano,}
Milan, Italy}
\IEEEauthorblockA{\orcidlink{0000-0002-9463-7146} \href{https://orcid.org/0000-0002-9463-7146}{0000-0002-9463-7146}}
\and
\IEEEauthorblockN{Davide Emilio Galli}
\IEEEauthorblockA{\textit{Università degli Studi di Milano} \\
Milan, Italy}
\IEEEauthorblockA{\orcidlink{0000-0002-1312-1181} \href{https://orcid.org/0000-0002-1312-1181}{0000-0002-1312-1181}}
\and
\IEEEauthorblockN{Enrico Prati}
\IEEEauthorblockA{\textit{CNR-IFN} \\
Milan, Italy}
\IEEEauthorblockA{\orcidlink{0000-0001-9839-202X} \href{https://orcid.org/0000-0001-9839-202X}{0000-0001-9839-202X}}
}

\maketitle

\begin{abstract}
Hybrid quantum-classical algorithms based on variational circuits are a promising approach to quantum machine
learning problems for near-term devices, but the selection of the variational ansatz is an open issue.
Recently, tensor network-inspired circuits have been proposed as a natural choice for such ansatz.
Their employment on binary classification tasks provided encouraging results.
However, their effectiveness on more difficult tasks is still unknown.

Here, we present numerical experiments on multi-class classifiers based on tree tensor network and multiscale entanglement renormalization ansatz circuits.

We conducted experiments on image classification with the MNIST dataset and on quantum phase recognition with the XXZ model by Cirq and TensorFlow Quantum.
In the former case, we reduced the number of classes to four to match the aimed output based on 2 qubits. The quantum data of the XXZ model consist of three classes of ground states prepared by a 
checkerboard circuit used for the ansatz of the variational quantum eigensolver, corresponding to three distinct quantum phases.
Test accuracy turned out to be 59\%-93\% and 82\%-96\% respectively, depending on the model architecture and on the type of preprocessing.
\end{abstract}


\section{Introduction}

\label{sec:introduction}
We implemented some multi-class variational classifiers based on tensor network circuits and we tested them on both classical and quantum data.

A lot of effort has been made recently to investigate exploitation of quantum computing \cite{rotta2017quantum,ferraro2020all} to the field of machine learning \cite{rocutto2021quantum,maronese2021continuous}.
Currently, a major research direction focuses on applications of noisy intermediate-scale quantum (NISQ) \cite{Preskill2018quantumcomputingin} devices i.e. near-term hardware characterized by a small number of noisy qubits.
Variational quantum classifiers \cite{Farhi_2018} trained in a hybrid quantum-classical setup are one of them.
They are classification algorithms that can natively process quantum data.
Therefore, they provide a strategy to manage future quantum datasets which may be classically intractable.
Nevertheless, they can also process classical datasets, once encoded in quantum states.
However, the choice of the variational \textit{ansatz} of such algorithms requires further investigations.

Tensor networks are mathematical objects used as variational ans\"atze to represent quantum states, initially developed in the field of condensed-matter physics \cite{OrusTensorNetworks, VerstraeteTensorNetworks, SchollwockTensorNetworks, Tensor_networks_for_complex_quantum_systems}.
However, their field of application turned out to be broader.
For example, tensor networks have been applied to machine learning tasks, e.g. supervised \cite{Stoudenmire2016, TN4ML, GTNC} and unsupervised \cite{Unsupervised_Generative_Modeling_Using_Matrix_Product_States} learning.
Then, they have been proposed as ans\"atze for variational quantum circuits applied to machine learning tasks \cite{TowardsQMLwithTNs}, in both discriminative \cite{Hierarchical_quantum_classifiers} and generative \cite{Generative_ML_with_TN_benchmark} learning.
Previous works \cite{TowardsQMLwithTNs, Hierarchical_quantum_classifiers} demonstrated promising results in binary classification problems, but the capabilities on more difficult tasks are still to be explored.

In this work we tested some circuit architectures based on tensor networks on multi-class classification tasks.
In particular, we trained the circuits for digit recognition with a subset of the MNIST dataset \cite{MNIST} and for quantum phase recognition with the 1D XXZ model \cite{Franchini, Ohanyan}.
In the former case, we compare the results with classical benchmarks.
We examined circuits based on both tree tensor networks (TTN) \cite{TTN} and multiscale entanglement renormalization ans\"atze (MERA) \cite{MERA}.

We adopted two different approaches to implement the multi-class setup, in the following referred to as amplitude decoding and qubit decoding with binary labels, respectively.
 The former retrieves the prediction from measurements in the computational basis.
The latter uses binary labels and retrieves the prediction by computing the expectation value of single-qubit observables in some readout qubits \cite{Stoudenmire2016, Qibo}. This research exploits circuits based on eight qubits.

In the case of digit recognition, the input data were compressed by dimensionality reduction methods.
In order to reduce the scale of the classification task, we filtered the dataset to include only four out of ten classes.
In the case of the XXZ model, we worked with a system of eight spins and with three different classes.

Test accuracy turned out to be 59\%-93\% for digit recognition and 82\%-96\% for quantum phase recognition, depending on the model architecture and on the type of preprocessing.

The document is organized as follows.
In Sec. \ref{sec:methods} we describe the implementation, while in Sec. \ref{sec:experiments} we present the numerical experiments.
Finally, in Sec. \ref{sec:conclusions} we draw the conclusions.

\section{Methods}
\label{sec:methods}
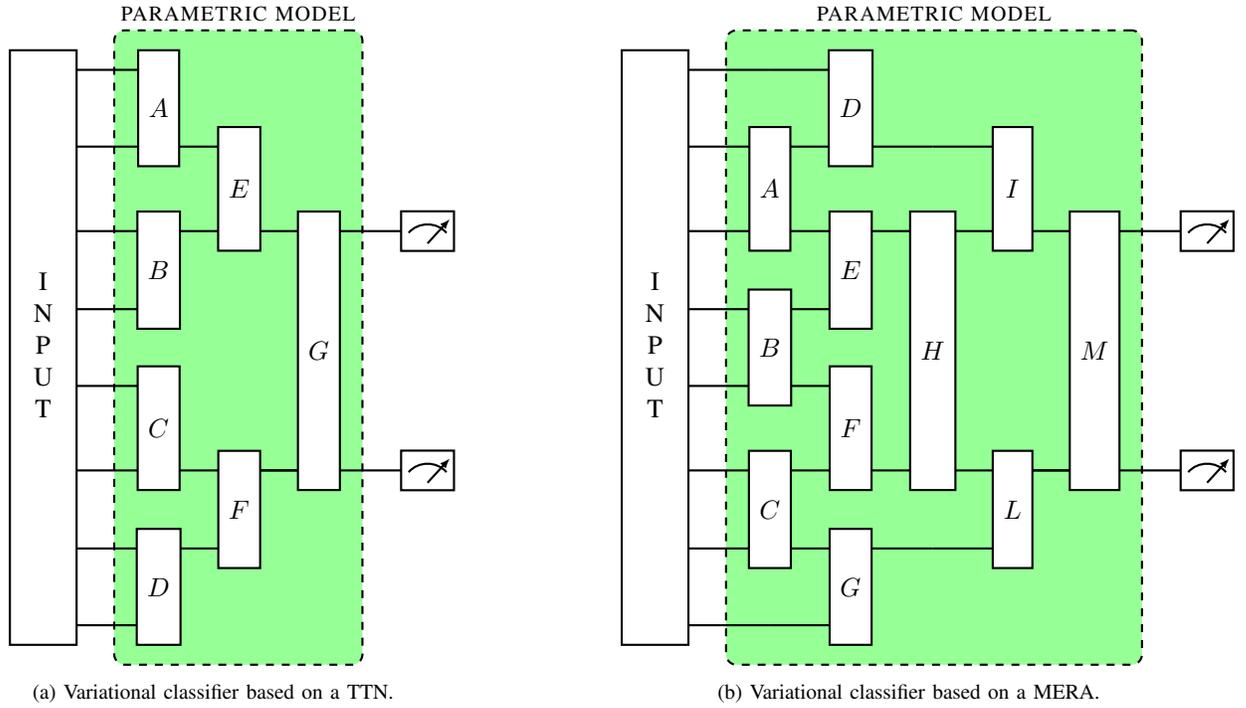
\begin{figure*}
    \centering
    \begin{subfigure}[b]{0.49\textwidth}
        \centering
        \begin{quantikz}
        & \gate[8, disable auto height, nwires={1,2,3,4,5,6,7,8}]{\begin{array}{c} \text{I} \\ \text{N} \\ \text{P} \\ \text{U} \\ \text{T} \end{array}} &[0.3cm] \gate[wires=2]{A} \gategroup[8,steps=3,style={dashed,rounded corners,fill=pqc, inner xsep=5pt},background]{{\sc parametric model}} & & &[0.3cm] \\
        &                       &                   & \gate[wires=2]{E} &     & \\
        &                       & \gate[wires=2]{B} &                   & \gate[4, nwires={2,3}]{G} & \meter{} \\
        &                       &                   &                   &     &\\
        &                       & \gate[wires=2]{C} &                   &     & \\
        &                       &                   & \gate[wires=2]{F} & \qw & \meter{} \\
        &                       & \gate[wires=2]{D} &                   &     & \\
        &                       &                   &                   &     &
        \end{quantikz}
        \caption{Variational classifier based on a TTN.}
        \label{circ:ttn}
    \end{subfigure}
    \hfill
    \begin{subfigure}[b]{0.49\textwidth}
        \centering
        \begin{quantikz}
        & \gate[wires=8, disable auto height, nwires={1,2,3,4,5,6,7,8}]{\begin{array}{c} \text{I} \\ \text{N} \\ \text{P} \\ \text{U} \\ \text{T} \end{array}} &[0.3cm] \qw \gategroup[8,steps=5,style={dashed,rounded corners,fill=pqc, inner xsep=5pt},background]{{\sc parametric model}} & \gate[wires=2]{D} &                                &                   &     &[0.3cm] \\
        &                                                 & \gate[wires=2]{A} &                   & \qw                            & \gate[wires=2]{I} & & \\
        &                                                 &                   & \gate[wires=2]{E} & \gate[wires=4,nwires={2,3}]{H} &                   & \gate[4, nwires={2,3}]{M} & \meter{} \\
        &                                                 & \gate[wires=2]{B} &                   &                                &                   &     &\\
        &                                                 &                   & \gate[wires=2]{F} &                                &                   &     & \\
        &                                                 & \gate[wires=2]{C} &                   &                                & \gate[wires=2]{L} & \qw & \meter{} \\
        &                                                 &                   & \gate[wires=2]{G} & \qw                            &                   &     & \\
        &                                                 & \qw               &                   &                                &                   &     &
        \end{quantikz}
        \caption{Variational classifier based on a MERA.}
        \label{circ:mera}
    \end{subfigure}
    \caption{Circuits used for the quantum classifiers.
             The nodes labelled by capital letters are parametric unitaries.
             In our experiments, we used those illustrated in Fig. \ref{circ:gates}.
             The input can be either an encoded classical state or a quantum state.
             The variational quantum circuit is applied to such input state.
             Then, some readout qubits are measured repeatedly to obtain the prediction.
             The number of readout qubits depends on the number of classification labels to predict and on the strategy used to decode the prediction from the final state of the circuit.
             Two readout qubits are used in the experiments of Sec. \ref{sec:experiments}.
             The representation of these circuits and of the following ones is made with \texttt{Quantikz} \cite{Quantikz}.}
    \label{circ:pqc}
\end{figure*}

The variational classifiers are built upon those proposed in \cite{TowardsQMLwithTNs, Hierarchical_quantum_classifiers} and modified to account for the multi-class embodiement. The core object of the algorithm is the variational ansatz i.e. a parametric circuit that takes the role of the model in a classical deep learning setup. We employed two different circuits, based on a TTN (Fig. \ref{circ:ttn}) and on a MERA (Fig. \ref{circ:mera}), respectively.
The inference step consists of applying the circuit to the input quantum state, then computing the expectation value of some observables on the final state.
In order to fit the input quantum state, classical data require an encoding.
Then, the learning process involves a hybrid quantum-classical iteration in which a batch of input data is fed into the circuit, a loss function is computed by comparing the predictions with the exact labels, and finally a classical optimization algorithm updates the parameters of the circuit in order to improve the performance of the classifier \cite{Farhi_2018}.
Besides the nature of the parametric model, this procedure is similar to the training process of classical deep learning \cite{Goodfellow-et-al-2016}.

In order to implement a multi-class setup, a method to decode the predictions from the final states is required.
As anticipated, we compare two different approaches, referred to as qubit decoding and amplitude decoding, respectively.

When the qubit decoding method is applied, we select $N$ readout qubits and we measure the expectation value of the operator $\frac{1}{2}\left( \mathbb{I} + \hat{\sigma}_z \right)$ for each readout qubit independently.
The operator $\frac{1}{2}\left( \mathbb{I} - \hat{\sigma}_z \right)$ would be an equivalent choice.
In this way, we obtain $N$ output values $\in [0, 1]$ that can be fed into a Softmax layer in order to transform them into a probability mass distribution for $N$ classes.
Instead, when the amplitude decoding method is applied, we select $n$ readout qubits and compute the probabilities of obtaining the $2^n$ elements of the computational basis by repeated measurements. Then, we obtain directly a probability mass distribution for $N = 2^n$ classes.
We still apply a Softmax layer, so that the result is normalized even if the number of classes is not a power of two.

The former strategy requires a number of readout qubits that scales linearly with the number of labels, while it scales as a logarithm for the latter.
Both strategies are compatible with one-hot encoded labels and with the categorical cross entropy loss function.

The linear scaling of the qubit decoding was troublesome for the scale of our circuits, therefore in our experiments we implemented a modified version of qubit decoding.
Here, the $N$ output values $\in [0, 1]$ are considered as the approximated prediction of a binary number.
Then, by converting the labels of the dataset to binary numbers, we can obtain a logarithmic scaling of the readout qubits with respect to the number of labels.
Notice that the binary labels are not equally distant (in terms of both the Hamming distance and the Euclidean distance), so a loss function like mean squared error will weight classification mistakes differently.

We report numerical experiments by using both qubit decoding with binary labels and amplitude decoding.
We obtained comparable results, without a strong evidence of advantage of one approach with respect to the other.
The details about the experiments and the results are discussed in the next Section.

\section{Experiments}
\label{sec:experiments}
The variational classifiers were built with \texttt{Cirq} \cite{cirq} and trained with \texttt{TensorFlow Quantum} \cite{TFQ}.
Data preprocessing and tests with other machine learning algorithms were performed with \texttt{Scikit-learn} \cite{scikit-learn} and \texttt{Keras} \cite{Keras}.

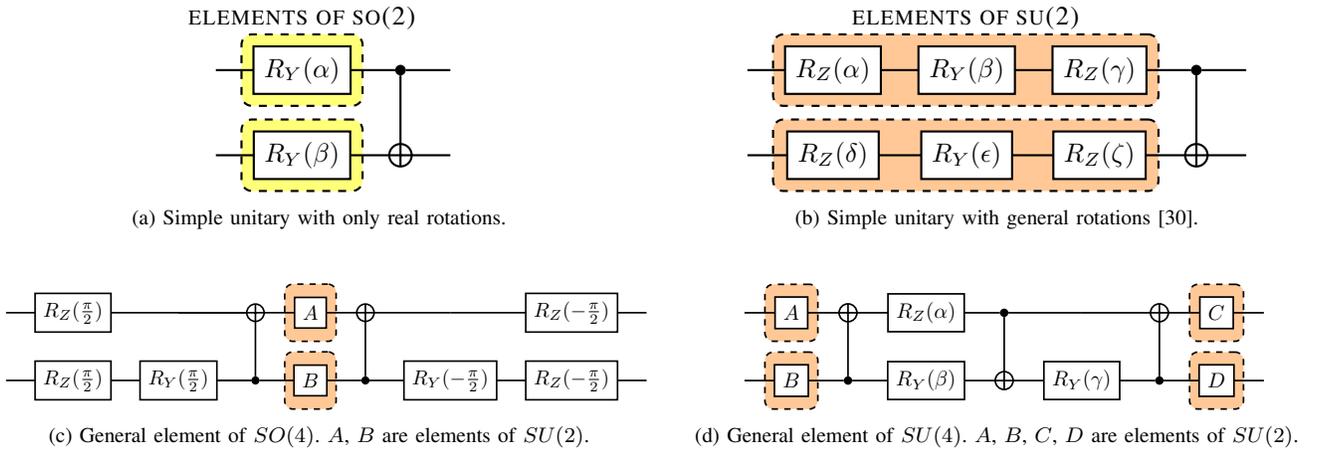
\begin{figure*}
    \centering
    \begin{subfigure}[b]{0.49\textwidth}
        \centering
        \begin{quantikz}
            \qw & \gate{R_Y(\alpha)} \gategroup[1,steps=1,style={dashed,rounded corners,fill=so2, inner xsep=1pt, inner ysep=1pt},background]{{\sc elements of so(2)}} & \ctrl{1} & \qw \\
            \qw & \gate{R_Y(\beta)} \gategroup[1,steps=1,style={dashed,rounded corners,fill=so2, inner xsep=1pt, inner ysep=1pt},background]{} & \targ{}  & \qw
        \end{quantikz}
        \caption{Simple unitary with only real rotations.}
        \label{circ:simple_real}
    \end{subfigure}
    \begin{subfigure}[b]{0.49\textwidth}
        \centering
        \begin{quantikz}
            \qw & \gate{R_Z(\alpha)} \gategroup[1,steps=3,style={dashed,rounded corners,fill=su2, inner xsep=1pt, inner ysep=1pt},background]{{\sc elements of su(2)}} & \gate{R_Y(\beta)} & \gate{R_Z(\gamma)} & \ctrl{1} & \qw \\
            \qw & \gate{R_Z(\delta)} \gategroup[1,steps=3,style={dashed,rounded corners,fill=su2, inner xsep=1pt, inner ysep=1pt},background]{} & \gate{R_Y(\epsilon)} & \gate{R_Z (\zeta)} & \targ{}  & \qw
        \end{quantikz}
        \caption{Simple unitary with general rotations \cite{Nielsen_Chuang}.}
        \label{circ:simple_complex}
    \end{subfigure}
    \begin{subfigure}[b]{0.49\textwidth}
        \centering
        \vspace{0.5cm}
        \resizebox{\textwidth}{!}{
        \begin{quantikz}
         & \gate{R_Z(\frac{\pi}{2})} & \qw                 & \targ{}   \qw & \gate{A} \gategroup[1,steps=1,style={dashed,rounded corners,fill=su2, inner xsep=1pt, inner ysep=1pt},background]{} & \targ{}   & \qw                  & \gate{R_Z(-\frac{\pi}{2})} & \qw \\
         & \gate{R_Z(\frac{\pi}{2})} & \gate{R_Y(\frac{\pi}{2})} & \ctrl{-1} \qw & \gate{B} \gategroup[1,steps=1,style={dashed,rounded corners,fill=su2, inner xsep=1pt, inner ysep=1pt},background]{} & \ctrl{-1} & \gate{R_Y(-\frac{\pi}{2})} & \gate{R_Z(-\frac{\pi}{2})} & \qw
        \end{quantikz}}
        \caption{General element of $SO(4)$. $A$, $B$ are elements of $SU(2)$.}
        \label{circ:so4}
    \end{subfigure}
    \begin{subfigure}[b]{0.49\textwidth}
        \centering
        \vspace{0.5cm}
        \resizebox{0.82\textwidth}{!}{
        \begin{quantikz}
         & \gate{A} \gategroup[1,steps=1,style={dashed,rounded corners,fill=su2, inner xsep=1pt, inner ysep=1pt},background]{} & \targ{}   & \gate{R_Z(\alpha)} & \ctrl{1} & \qw                & \targ{}   & \gate{C} \gategroup[1,steps=1,style={dashed,rounded corners,fill=su2, inner xsep=1pt, inner ysep=1pt},background]{} & \qw \\
         & \gate{B} \gategroup[1,steps=1,style={dashed,rounded corners,fill=su2, inner xsep=1pt, inner ysep=1pt},background]{} & \ctrl{-1} & \gate{R_Y(\beta)}  & \targ{}  & \gate{R_Y(\gamma)} & \ctrl{-1} & \gate{D} \gategroup[1,steps=1,style={dashed,rounded corners,fill=su2, inner xsep=1pt, inner ysep=1pt},background]{} & \qw
        \end{quantikz}}
        \caption{General element of $SU(4)$. $A$, $B$, $C$, $D$ are elements of $SU(2)$.}
        \label{circ:su4}
    \end{subfigure}
    \caption{Quantum circuits that implement the nodes of the tensor networks of our experiments.
             Circuit \ref{circ:simple_real} and Circuit \ref{circ:simple_complex} are taken from \cite{Hierarchical_quantum_classifiers},
             while the other two are derived in \cite{Two-qubits_gates}.
             For the first two, the CNOT gates may be reversed to follow the causal structure of the network e.g. gates $B$, $D$ and $F$ of Fig. \ref{circ:ttn}.}
    \label{circ:gates}
\end{figure*}

The choice of the parametric unitaries that implement the nodes of the networks was inspired by \cite{Hierarchical_quantum_classifiers}.
In particular, we used two simple unitaries that use only one CNOT, plus a general $SO(4)$ gate and a general $SU(4)$ gate.
We compiled the unitaries with CNOTs and parametrized single-qubit rotations.
The corresponding circuits are described in Fig. \ref{circ:gates}.
When we used one of the two simple unitaries, we added single-qubit rotations to the readout qubits at the end of the circuit.
Such single-qubit rotations were the same of those used in the unitaries.

In Sec. \ref{subsec:mnist} we describe the tests that we conducted on the task of digit recognition with the MNIST dataset, while in Sec. \ref{subsec:xxz} we describe those aimed to quantum phase recognition referring to XXZ model.

\subsection{Digit recognition with MNIST}
\label{subsec:mnist}
\begin{figure}[tp]
    \centering
    \includegraphics[width=\columnwidth]{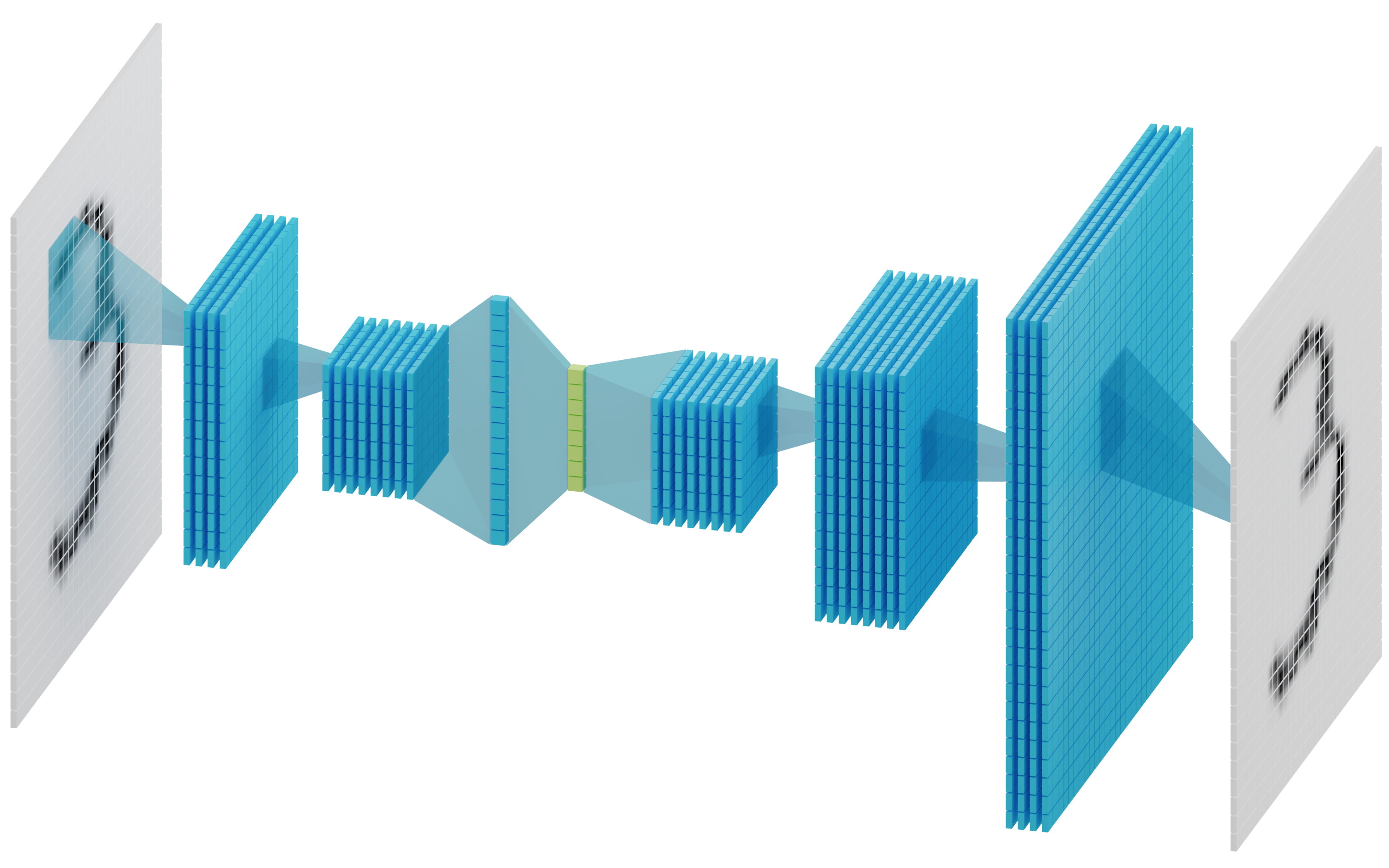}
    \caption{Illustration of a convolutional autoencoder.
             The input data are transformed by convolutional and fully connected layers into a representation in a latent space (shown in green).
             Then, the representation is transformed again by a neural network to reconstruct the input.}
    \label{fig:conv_autoencoder_illustration}
\end{figure}

We apply the classifiers to the task of digit recognition on the MNIST dataset \cite{MNIST}.
The dataset consists of 70000 grayscale images of 28x28 pixels depicting handwritten digits from zero to nine, subdivided between training and test set with a ratio 6:1.
In order to use circuits with a small number of qubits, we reduced the dimensionality of the data from 784 to 8 by using two alternative methods. First, we applied Principal Component Analysis (PCA) \cite{PCA_review, Hierarchical_quantum_classifiers}, by keeping only the components with higher variance.
Second, we performed the dimensionality reduction with a convolutional autoencoder \cite{Geron} to see whether a deep learning algorithm could improve the performance of the classifiers.
Autoencoders are machine learning algorithms based on neural networks that are able to learn representations of the input data.
Autoencoders are made of two submodels: an encoder that maps the data into a latent space and a decoder that reconstructs the input data from the latent space.
The autoencoder is trained by optimizing the quality of the reconstruction i.e. a distance between the original example and its reconstruction.
Once that the model is trained, the encoder can be used to encode the examples into the latent space, which is supposed to have a lower dimensionality than the original input space, so that a dimensionality reduction is achieved.
An illustration of a convolutional autoencoder is presented in Fig. \ref{fig:conv_autoencoder_illustration}, while a description of the specific configuration we used is given in Table \ref{tab:hyper_conf}.
\begin{table}[tp]
    \centering
    \renewcommand{\arraystretch}{1.3}
    \caption{Architecture of the autoencoder used in Sec. \ref{subsec:mnist}, adapted from the code examples of \texttt{Keras} \cite{Keras}.}
    \label{tab:hyper_conf}
    \begin{tabular}{c c c}
        \toprule
        Layer & Output shape & Activation \\
        \midrule
        Input & 28x28x1 & \\
        Convolutional layer & 14x14x32 & ReLU \\
        Convolutional layer & 7x7x64  & ReLU \\
        Fully connected layer & 100 & ReLU \\
        Fully connected layer & 8 & ReLU \\
        \midrule
        Fully connected layer & 7x7x64 & ReLU \\
        Transposed conv. layer & 14x14x64 & ReLU \\
        Transposed conv. layer & 28x28x32 & ReLU \\
        Transposed conv. layer & 28x28x1 & Sigmoid \\
        \midrule
        \multicolumn{3}{c}{Configuration of convolutional layers}\\
        Same padding & \multicolumn{2}{c}{Filter size: 3x3} \\
        \midrule
        Optimizer: Adam \cite{Adam} & Epochs: 100 & Batch size: 128 \\
        \multicolumn{3}{c}{Loss function: Binary Cross-entropy} \\
        \bottomrule
    \end{tabular}
\end{table}

We encoded the classical data in quantum circuits by using the qubit encoding i.e. by constructing a factorized state in which each predictor is encoded in the amplitude of a single-qubit wavefunction.
This type of encoding requires a qubit for each predictor.
In particular, we used the approach introduced in \cite{Stoudenmire2016}.
Let $X$ be the design matrix of the training set, with examples in the first index and predictors in the second index. 
At first, we scale the predictors to be $\in [0, 1]$ within the training set.
Then, we encode each example $X_{i,:}$ in a quantum state using $R_Y$ rotations:
\begin{equation}
    \ket{\psi_i} = R_Y(\pi x_{i1}) \otimes R_Y(\pi x_{i2}) \otimes ... \otimes R_Y(\pi x_{i8}) \ket{0}^{\otimes 8}
\end{equation}
where $x_{ij}$ are the elements of the design matrix $X$.
The encoding requires a number of one-qubit rotations that scales linearly with the number of predictors.

The preprocessing transformations (dimensionality reduction and scaling) are applied to both the training and the test set, but the parameters of these transformations are computed without using the elements of the test set. 

The circuits of Fig. \ref{circ:pqc} are TTN and MERA circuits with bond dimension two.
They natively support at most two readout qubits at the last node of the circuit.
Therefore, we reduced the number of different labels of the dataset from ten to four.
In particular, we kept the digits `0', `1', `2' and `3' so the required number of readout qubits is two for both amplitude decoding and qubit decoding with binary labels.

Concerning the training, we use the Adam \cite{Adam} optimizer. The gradient is computed with the default differentiator of \texttt{TensorFlow Quantum}.
We used batches of 20 examples, epochs of 10 batches and default learning rate (0.001).
Moreover, we further split the training set between training and validation set (ratio 11:1) to use the Early Stopping method \cite{Goodfellow-et-al-2016}.
We set the patience of the Early Stopping to 100, monitoring validation accuracy.
The maximum number of epochs was set to 1000.
Such settings were not fine tuned.
Instead, we followed another configuration that worked well in the binary case \cite{Hierarchical_quantum_classifiers}.
Further hyperparameter tuning may improve the results.
We tested both the qubit decoding with binary labels and amplitude decoding approaches presented in Sec. \ref{sec:methods}, respectively, with mean squared error and categorical cross entropy as loss functions.
The two approaches share the same parametric quantum circuit, with the only difference being the interpretation of the measurements.

The results are presented in Table \ref{table:results_mnist}.
\begin{table*}[tp]
    \centering
    \renewcommand{\arraystretch}{1.3}
    \caption{Experimental results with the MNIST dataset}
    \label{table:results_mnist}
    \begin{minipage}{\textwidth}
    \begin{tabular}{c c c c c c}
    	\toprule
    	\multirow{3}{*}{Classifier} & \multirow{3}{*}{Unitaries} & \multicolumn{4}{c}{Test accuracy$^\mathrm{a}$ (\%)} \\
        & & \multicolumn{2}{c}{PCA} & \multicolumn{2}{c}{Convolutional autoencoder} \\
        & & Qubit decoding w/ binary labels & Amplitude decoding & Qubit decoding w/ binary labels & Amplitude decoding \\
    	\midrule
        TTN  & Simple  $SO(4)$ (\ref{circ:simple_real})    & $64 \pm 5$     & $59 \pm 10$    & $76.1 \pm 1.2$ & $76 \pm 3$ \\
        TTN  & Simple  $SU(4)$ (\ref{circ:simple_complex}) & $64 \pm 7$     & $69 \pm 10$    & $75 \pm 4$     & $74 \pm 6$\\
        TTN  & General $SO(4)$ (\ref{circ:so4})            & $77 \pm 7$     & $69 \pm 5$     & $81 \pm 4$     & $80 \pm 4$\\
        TTN  & General $SU(4)$ (\ref{circ:su4})            & $81.0 \pm 1.6$ & $68 \pm 10$    & $83 \pm 3$     & $82 \pm 5$\\
        MERA & General $SO(4)$ (\ref{circ:so4})            & $82.8 \pm 1.4$ & $81 \pm 7$     & $90 \pm 3$     & $91 \pm 3$ \\
        MERA & General $SU(4)$ (\ref{circ:su4})            & $84.8 \pm 1.6$ & $85.0 \pm 1.9$ & $91.3 \pm 1.0$ & $93 \pm 2$\\
        \midrule
        \multicolumn{2}{c}{Logistic regression}            & \multicolumn{2}{c}{94} & \multicolumn{2}{c}{98} \\
        \multicolumn{2}{c}{Neural network (186 parameters$^\mathrm{b}$)}    & \multicolumn{2}{c}{95} & \multicolumn{2}{c}{99} \\
    	\bottomrule
    \end{tabular}
    \footnotetext{$^\mathrm{a}$The results of the quantum classifiers are averaged over five trials, with different random seeds. The measurement of uncertainty is the standard deviation.}
    \footnotetext{$^\mathrm{b}$The architecture of the neural network was chosen in order to have a number of free parameters which is similar to that of the most expressive quantum circuit we used (that is 165).}
    \end{minipage}
\end{table*}
The quantum classifiers achieved solid results, even though they were not able to beat the classical algorithms we used for reference.
In addition, the dimensionality reduction carried out with the convolutional autoencoder outperforms that given by the PCA, in the sense that it allows the classifiers to learn better.
Notice that the initialization and training process of the autoencoder is not deterministic, so that different trained instances of the same model can give different results.
The results of Table \ref{table:results_mnist} involve a specific instance.
Another instance showed worse results, but only for qubit decoding with binary labels.

\subsection{Quantum phase recognition with the XXZ model}
\label{subsec:xxz}
Now we turn to the application of the classifiers to the task of quantum phase recognition, adopting the framework proposed in \cite{ML_phase_transitions_with_a_quantum_processor}.

Let us first define the task.
Consider a quantum system with a Hamiltonian $H(\Delta)$ which presents different phases at different values of $\Delta$.
We can use the Variational Quantum Eigensolver algorithm (VQE) \cite{VQE} to obtain a circuit that prepares a state $\ket{\psi(\Delta)}$ which approximates the ground state of $H(\Delta)$. 
Then, we can build a dataset of ground state circuits $\{ \ket{\psi(\Delta)}, y(\Delta)\}_{\Delta \in \Lambda}$, where $\ket{\psi(\Delta)}$ is a variational approximation to the ground state wave function of $H(\Delta)$, $y(\Delta)$ is a label for the quantum phase of the system and $\Lambda$ is the set of different values of the parameter $\Delta$ with which we want to generate the dataset.

We selected the one-dimensional XXZ model as our benchmark system. It has the following Hamiltonian, at least in absence of an external magnetic field:
\begin{equation}
    H(\Delta) = J \sum_{i=1}^N \left[ \sigma_i^x \sigma_{i+1}^x + \sigma_i^y \sigma_{i+1}^y + \Delta \sigma_i^z \sigma_{i+1}^z \right]
\end{equation}
Let us consider a spin chain of length $N=8$ with periodic boundary conditions i.e. $\sigma_{i+N} = \sigma_i$.
Let $J=1$ so that the antiferromagnetic order is favoured along the $x-y$ plane.
The parameter $\Delta$ regulates the intensity of the $z$-axis anisotropy with respect to the planar $x-y$ term and discriminates between axial ($|\Delta| > 1$) and planar ($|\Delta| < 1$) regimes.
Regarding the axial regimes, for $\Delta > 1$ we have a gapped antiferromagnet along the $z$-axis, while for $\Delta < -1$ we have a gapped \mbox{$z$-axis} ferromagnet.
For $|\Delta| < 1$ the system exhibits a gapless paramagnetic phase \cite{Franchini, Ohanyan}.

\begin{figure}[tp]
    \centering
    \includegraphics[width=0.64\columnwidth]{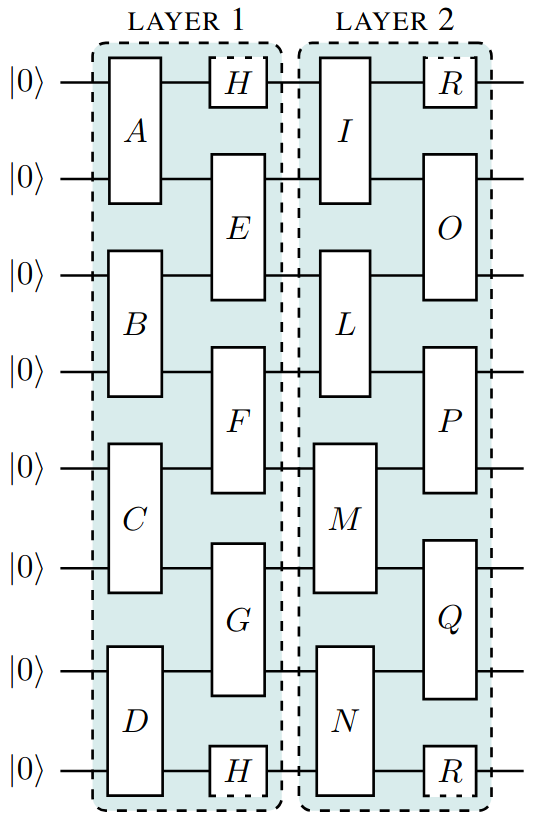}
    \caption{The checkerboard circuit used for the ground state ansatz of the VQE.
             The number of layers was chosen to obtain a good compromise between the quality of the results and the computational cost.
             The gates labelled by capital letters are general $SU(4)$ gates (see Fig. \ref{circ:su4}).
             We did not implement parameter sharing because initial experiments showed worse results, at least with our configuration.}
    \label{circ:checkerboard}
\end{figure}

We implemented the VQE by using \texttt{TensorFlow Quantum} with the Adam optimizer \cite{Adam}.
In particular, the optimization was performed by minimizing the mean squared distance between the energy of the variational state and a target energy, which was chosen to be lower than the exact ground state energy.
We partially reused the code of \cite{ML_phase_transitions_with_a_quantum_processor}.
The variational ansatz was a checkerboard tensor network made with general $SU(4)$ gates.
It is illustrated in Fig. \ref{circ:checkerboard}.

We generated 1000 ground states for $\Delta \in [-2,2]$.
Then, we compared the ground state energies obtained with the VQE with those computed by exact diagonalization, resulting in relative errors up to 4-5\%.
While our results are probably sub-optimal, we found out that it was enough to our goal.

The dataset consists of quantum data that can be fed directly to the classifiers, without any data encoding process.
In practice, the examples consist of quantum circuits that can be readily added in front of the classifiers. 
The number of required operations depends on the complexity of the variational ansatz of the VQE.

We split the dataset in training and test set with ratio 2:1, then we trained the classifiers on the training set and we evaluated their performance on the test set.
Concerning the training, we used the Adam \cite{Adam} optimizer and the gradient was computed with the default differentiator of \texttt{TensorFlow Quantum}.
We used batches of 8 examples and default learning rate (0.001).
Moreover, we further split the training set in training and validation set (ratio 11:1) to use the Early Stopping method \cite{Goodfellow-et-al-2016}.
We set the patience of the Early Stopping to 250, monitoring validation accuracy.
The maximum number of epochs was set to 1000.
We tried both the qubit decoding with binary labels and amplitude decoding approaches presented in Sec. \ref{sec:methods}, respectively with mean squared error and categorical cross entropy as loss functions.

The results are presented in Table \ref{table:results_xxz}.
Overall, the classifiers were quite accurate.
Those implemented with qubit decoding with binary labels achieved a test accuracy of 87\%-96\%, depending on the model complexity.
The results for the amplitude decoding were slightly worse (82\%-90\%).

\begin{table}[tp]
    \centering
    \renewcommand{\arraystretch}{1.3}
    \caption{Experimental results with the XXZ dataset}
    \begin{minipage}{\columnwidth}
    \centering
    \begin{tabular}{c c c c}
    \toprule
        \multirow{2}{*}{Classifier} & \multirow{2}{*}{Unitaries} & \multicolumn{2}{c}{Test accuracy$^\mathrm{c}$ (\%)} \\
                                    &                            & Qubit dec. & Amplitude dec. \\
    \midrule
        TTN  & Simple  $SO(4)$ (\ref{circ:simple_real})    & $87 \pm 4$     & $82 \pm 5$ \\
        TTN  & Simple  $SU(4)$ (\ref{circ:simple_complex}) & $91 \pm 7$     & $82 \pm 10$ \\
        TTN  & General $SO(4)$ (\ref{circ:so4})            & $94 \pm 2$     & $82 \pm 4$\\
        TTN  & General $SU(4)$ (\ref{circ:su4})            & $91 \pm 4$     & $86 \pm 6$\\
        MERA & General $SO(4)$ (\ref{circ:so4})            & $95.6 \pm 0.7$ & $90 \pm 7$\\
        MERA & General $SU(4)$ (\ref{circ:su4})            & $96.0 \pm 0.5$ & $90 \pm 5$\\
    \bottomrule
    \end{tabular}
    \footnotetext{$^\mathrm{c}$The results are averaged over five trials, with different random seeds.
             The measurement of uncertainty is the standard deviation.}
    \end{minipage}
    \label{table:results_xxz}
\end{table}

\section{Conclusions}
\label{sec:conclusions}
We implemented variational quantum classifiers based on TTNs and MERAs and we tested them on multi-class classification tasks, namely digit recognition with the MNIST dataset and quantum phase recognition with the XXZ model.
We obtained a classification accuracy of 59\%-93\% and 82\%-96\% respectively, depending on the class of tensor network, the choice of the unitaries, the choice of the decoding method and the type of preprocessing. MERAs perform  better than TTNs.
The tasks considered in this work consists of discrimination up to four different classes.
Such number of classes is naturally manageable by networks of bond dimension two.
The results with four classes are promising as networks with higher bond dimension may be able to manage more complicated datasets, provided that an efficient implementation method is available. In particular, such networks require the implementation of multi-qubit parametric unitaries.

For quantum data like the dataset of XXZ ground states considered above, the elements of the dataset are quantum states.
Providing such states as input to a classical machine learning algorithm may be intractable, while the quantum classifiers presented here can process such states natively.
The computations underlying this work have been simulated on classical hardware.
Future advances of quantum processors and quantum algorithms may provide quantum datasets which are not classically tractable, so that development of quantum classifiers may turn out crucial.

\section*{Acknowledgment}
\label{sec:acknowledgment}
Computational resources were provided by INDACO Platform, which is a project of High Performance Computing at the University of Milan.

\renewcommand*{\UrlFont}{\rmfamily}
\printbibliography

\end{document}